\documentclass{emulateapj}

\shortauthors{Parker et al.}
\shorttitle{Lensing Mass-to-Light Ratios of Groups}

\begin{document}

\def \farcs{\hbox{$.\!\!^{\prime\prime}$}}
\def \farcm{\hbox{$.\!\!^{\prime}$}}

\title{Mass-to-Light Ratios of Galaxy Groups from Weak Lensing}

\shorttitle{Lensing Mass-to-Light Ratios of Groups}

\author{Laura C. Parker\altaffilmark{1} \altaffilmark{2} and Michael J. Hudson\altaffilmark{1}}
\affil{Department of Physics, University of Waterloo, Waterloo, ON, N2L 3G1}
\altaffiltext{1}{Visiting Astronomer, Kitt Peak National Observatory,
  National Optical Astronomy Observatory, which is operated by the
  Association of Universities for Research in Astronomy, Inc. (AURA)
  under cooperative agreement with the National Science Foundation}
\altaffiltext{2}{Based on observations obtained at the Canada-France-Hawaii
        Telescope (CFHT) which is operated by the National Research
        Council of Canada, the Institut National des Science de
        l'Univers of the Centre National de la Recherche Scientifique of
        France,
        and the University of Hawaii.}

\author{R.G. Carlberg}
\affil{Department of Astronomy \& Astrophysics,University of Toronto, Toronto, ON,  M5S 3H8}

\and

\author{Henk Hoekstra}
\affil{Department of Physics and Astronomy, University of Victoria, Victoria, BC,  V8W 2Y2}

\begin{abstract}

We present the findings of our weak lensing study of a sample of 116
CNOC2 galaxy groups. The lensing signal is used to estimate the
mass-to-light ratio of these galaxy groups. The best fit isothermal sphere model to our lensing data has an Einstein radius of
0{\farcs}88$\pm$0{\farcs}12, which corresponds to a shear-weighted velocity
dispersion of 245$\pm$18 km s$^{-1}$. The mean mass-to-light ratio within 1 h$^{-1}$ Mpc is 185$\pm$28 hM$_\odot$/L$_{B\odot}$ and is independent of radius from the group center.

The signal-to-noise ratio of the shear measurement is sufficient to split the sample into subsets of ``poor'' and ``rich'' galaxy groups. The poor galaxy groups were
found to have an average velocity dispersion of
193$\pm38$ km s$^{-1}$ and a mass-to-light ratio of 134$\pm$26
hM$_{\odot}$/L$_{B\odot}$, while the rich
galaxy groups have a velocity dispersion of 270$\pm39$ km s$^{-1}$ and a
mass-to-light ratio of 278$\pm$42 hM$_{\odot}$/L$_{B\odot}$, similar to the mass-to-light ratio of clusters. This steep increase in
the mass-to-light ratio as a function of mass, suggests that the mass scale of $\sim 10^{13}$M$_{\odot}$ is where the transition between the actively star-forming field environment and the passively-evolving cluster environment occurs. This is the first such
detection from weak lensing.

\end{abstract}

\keywords{gravitational lensing, galaxy groups, dark matter,
 mass-to-light ratios, galaxy halos}

\clearpage

\section{Introduction}

Galaxy groups dominate the overall mass and luminosity densities of
the Universe yet their properties are poorly understood in comparison to individual galaxies
or rich galaxy clusters. Until now galaxy groups have not been extensively used for
cosmology largely because they are notoriously difficult to identify due to
their small contrast with the field. However, with
large redshift surveys it is now possible to identify substantial
samples of galaxy groups.

A sample of roughly 200 galaxy groups was identified in the Canadian
Network for Observational Cosmology 2 (CNOC2)
redshift survey using an iterative friends-of-friends algorithm
(Carlberg et al. 2001). The dynamical analysis of these groups indicated a
rising mass-to-light ratio with radius. This suggests that groups are the scale where
segregation begins to occur between mass and light. This effect could
be due to dynamical friction, or to a large core radius which
could indicate that dark matter has different properties from
``standard'' collisionless cold dark matter (CDM). The cores of
galaxies and clusters appear to be less cuspy than expected, which has
prompted theoretical work in alternative dark matter models (see
discussion in Governato et al. 2001).

 The dark matter density profile has yet to be measured for galaxy groups.
 Dynamical studies of groups are difficult because kinematic
 information is known for very few galaxies, and because equilibrium
 assumptions might not be valid. Furthermore, these difficulties increase at large
 radii from the group
 center. Weak gravitational lensing has proven invaluable in the
 analysis of single massive objects such as galaxy clusters (Hoekstra
 et al. 1998; Mellier 1999) as well as in the statistical studies of
 individual galaxies (Brainerd, Blandford \& Smail, 1996; Hudson et al. 1998; Fischer et al. 2000; Sheldon et al. 
2001; Hoekstra et al. 2004). To date there has been only one weak lensing measurement of galaxy groups (Hoekstra et al. 2001) using a small subsample of the total CNOC2
 galaxy group catalog.

Assuming that the dark matter halos of groups are well described by an isothermal sphere, we expect a tangential shear signal as follows
\begin{equation}
\gamma_T=\frac{\theta_E}{2\theta}=\frac{4\pi\sigma^2}{c^2}\frac{D_{LS}}{D_S}
\end{equation}
where $\sigma$ is the velocity dispersion of the halo, and D$_S$ and D$_{LS}$ are the angular diameter distances to the source and between lens and source, respectively.

 The intent of this paper is to present the results of our weak lensing
 study of CNOC2 galaxy groups and to compare these results with those found
 from the dynamical measurements (Carlberg et al. 2001) and the
 weak lensing detection of Hoekstra et al. (2001). We will also present
 the results when the sample of groups is split into two samples of
 ``rich galaxy groups'' and ``poor galaxy groups'', divided by
 velocity dispersion. A second
 paper will follow with the results of galaxy-galaxy lensing in these
 fields and a maximum likelihood analysis of the shear.

\section{Data}

\subsection{CNOC2 Groups}

Our galaxy group catalogs were generated using a friends-of-friends
algorithm with the CNOC2 redshift survey data (Yee et al. 2000; Carlberg et al. 2001). The CNOC2 area contains 4
fields well-spaced in right ascension and was intended to better
understand the properties of field galaxies. The CNOC2 galaxy
sample contains 6200 galaxies with redshifts to z of 0.7. From this
galaxy catalog a sample of 192 galaxy groups was identified. The
average number of galaxies identified in each group is $\sim$4 and the
groups have a
median redshift of 0.33. The groups have a median dynamically
determined velocity dispersion of 190 km s$^{-1}$.

\subsection{Observations}

For this project we
observed the 4 central patches of the CNOC2 fields, where most of the
galaxy groups are located. The observations were carried out mostly at
the Canada-France-Hawaii Telescope with 2 additional nights at
the Kitt Peak National Observatory Mayall 4-m Telescope. The fields were observed in B,V,$R_c$, and
$I_c$. Deep exposures ($\sim$4 hours) were taken in the $R_c$ and $I_c$ bands, which were
used for the lensing measurements. The characteristics of the data
obtained are outlined in Table 1.

\begin{table*}
\begin{center}
\caption{Field Information.\label{tbl-1}}
\begin{tabular}{cccccccc}
\tableline
Field & RA & DEC & Telescope & Area & Source Density & Median  & No. of \\

& (2000) & (2000) & & sq.arcmin &(No./sq arcmin) & Seeing & Groups\\

\tableline
0223 &36.51992 & 0.3116 & KPNO & 1120 & 30 & 1.1 & 23 \\
0920 &140.95504 & 37.0861 & CFHT & 1100 & 45 & 0.9 & 40 \\
1447 &222.4096 & 9.13883 & CFHT & 1220 & 32 & 0.8 & 29 \\
2148 &327.8317 & -5.5586 & CFHT & 1125 & 44 & 0.8 & 25 \\
\tableline
\end{tabular}
\end{center}
\end{table*}

\subsection{Reduction/Stacking}

Gravitational lensing is usually limited
by systematics and it is important to ensure no spurious shear is
introduced in the stacking procedure. This can be achieved by
carefully monitoring the astrometry over each input image that enters
the stack. Wide-field cameras in use today have larger distortions
than earlier, smaller CCD cameras. This distortion must be properly
mapped and corrected in order to ensure no artificial source of shear
is imported during the stacking process. Note, however, that group lensing is less affected by systematics than cosmic
shear studies (the weak lensing signal from the large scale structure
in the Universe). This is due to the random orientation of the
galaxy-group pairs across the field, as opposed to looking for a
preferred orientation as cosmic shear studies do. Cosmic shear measurements use the patterns in the large-scale distortion field of background sources to map out the matter distribution in the Universe. This signal is tiny and more susceptible to systematics than galaxy-galaxy or group-galaxy lensing where the shear signal is averaged in radial bins around each lens. In this analysis the image reduction and stacking was carried out using the IRAF mosaic package
mscred (Valdes, F.G., 1997).

\subsection{Object Detection and Shape Parameters}

Object catalogs were extracted from our stacked images using the imcat
software, an implementation of the Kaiser, Squires and Broadhurst
(1995, hereafter KSB) method. This software is
optimized for measuring the shapes of faint sources.
The object detection algorithm
works by smoothing the images using different sized filters and then
detecting the ``peaks'' which are then added to the source
catalog. For each detected object weighted quadrupole moments were
measured and the resulting polarizations were calculated:
\begin{equation}
e_1=\frac{I_{11}-I_{22}}{I_{11}+I_{22}}  \mbox{ and }  e_2=\frac{2I_{12}}{I_{11}+I_{2}}
\end{equation}

The polarization measurements need to be corrected for the effects of
seeing, camera distortion and PSF anisotropy. These corrections
to correct for these concerns have been discussed in KSB and Luppino \&
Kaiser (1997) with some improvements made by Hoekstra et al. (1998 and
2000). The techniques work well for ground-based data where the PSF is
stable and not very anisotropic, and where the fields contain many
stars which are used in the correction algorithms. The source
catalogs are trimmed so that all stars are removed. The stars can easily be located by comparing magnitude and half light radius. We kept only those objects for which the half light radii were greater than 1.2 times the stellar PSF, thus ensuring the contamination from stars in our source catalog is very small. The limiting magnitude of our images is approximately 25 in R$_c$.

\section{Analysis}

\subsection{Weak Lensing Measurement}

In the weak lensing analysis we used a source catalog of
approximately 150 000 objects ($\sim$40 per sq arcminute) and a galaxy group catalog containing the
116 CNOC2 galaxy group centers that were within the area we
observed. The faint members of the galaxy groups are included in
the source catalog, but as shown by Hoekstra et al. (2001), this does
not contaminate the final result. This is indicated by the fact that the number density of faint galaxies does not increase significantly towards the group center, thus faint group members are do not influence the final shear measurement.

The source density of background objects is not sufficiently high to extract a
signal from individual galaxy groups, except for the most massive
groups, and so the galaxy groups must be
stacked and the weak lensing signal measured around the stacked
groups. The source galaxies around the stacked galaxy group were
divided into radial bins and the average distortion
was calculated in each bin. The component of the average distortion tangential to
the group center is the weak lensing signal and is displayed in
Figure 1a. The tangential shear is plotted in physical bins (units of h$^{-1}$Mpc) since the redshift of each galaxy group is precisely known from the CNOC2 redshift survey. Using equation (1), the best fit isothermal sphere to the average tangential shear profile
yielded
an Einstein radius of 0{\farcs}88$\pm$0{\farcs}13.

We can alternatively fit the tangential shear data with a Navarro, Frenk and White (NFW) dark matter profile (Navarro, Frenk \& White, 1996). This density profile, which has been observed to fit mass distributions well over a wide range of scales, is given by 
\begin{equation}
\rho(r)=\frac{\delta_c\rho_c}{(r/r_s)(1+r/r_s)^2}
\end{equation}
 where $\rho_c$ is the critical density for closure of the Universe. The scale radius, $r_s$, is defined as $r_{200}/c$ where c is the dimensionless concentration parameter, and $\delta_c$ is the characteristic over-density of the halo. The tangential shear signal $\gamma_T$ as a function of radius $\theta$ for a NFW halo is given by (Wright \& Brainerd, 2000). 

We use the tangential shear data to do a one-parameter fit to the NFW profile, while assuming a reasonable value for the concentration parameter, c. Based on the high resolution numerical simulations of Bullock et al. (2001), a concentration parameter of $\sim$10 was used as an estimate for galaxy group scales. The best fit NFW profile can be seen as the dashed line in Fig 1a. Over the scales where we can measure the weak lensing signal the NFW profile and the isothermal sphere are very similar and are both good fits to the data. The similarity of these two models at the galaxy group mass scale is expected from models (Wright \& Brainerd, 2000).

A common systematic test employed in gravitational lensing is to
measure the signal when the phase of the distortion is increased by
$\pi/2$. If the measured tangential distortion is due to gravitational
lensing the rotated signal should be consistent with 0 as is shown in 
Figure 1b. In addition to the standard systematic test, we also
measured the signal around random points in the
field. This test yielded no signal which indicates the results
plotted in Figure 1a are indeed due to gravitational lensing by the groups.

\begin{figure}[h]
\epsscale{1.0}
\plotone{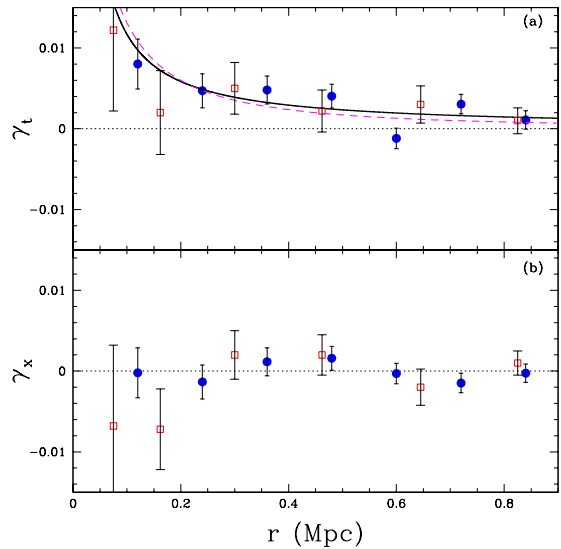}
\caption{(a)The ensemble averaged tangential shear as a function of radius
  around a sample of CNOC2 galaxy group centers from Carlberg et al
  (2000).The best fit isothermal sphere, shown with the solid line, yields an Einstein radius of
  0{\farcs}88$\pm$0{\farcs}13 corresponding to a velocity dispersion of 245$\pm$18 km s$^{-1}$. The best fit NFW profile is shown with the dashed curve. (b) The signal when the phase is rotated by
  $\pi/2$. No signal is present as expected if the signal in (a) is
  due to gravitational lensing. The results of Hoekstra et al. (2001)
  are shown with the open squares while the results of this paper are shown will filled circles. There is good agreement between
  the two measurements but the result here is a much more significant
  detection}

\end{figure}

\subsection{Velocity Dispersion}

In order to relate our estimate of the Einstein radius to the
average velocity dispersion of the groups, the redshift distribution of the
background sources must be understood. The strength of the
gravitational lensing signal as a function of redshift is
characterized by the parameter $\beta$ which is defined as
$\beta$=max$[0,D_{LS}/D_S]$ where $D_{LS}$ is the angular diameter
distance between the lens and the source and $D_S$ is the angular
diameter distance from the observer to the source. $\beta$ was calculated for each group-source pair based on the known spectroscopic redshift of the group and the estimated redshift of the source. The source redshift estimate was based on the observed R$_c$ magnitude and the method outlined in Brainerd, Blanford and Smail
(1996). We find a value of $\beta$=0.49. This yields an ensemble averaged group velocity dispersion
 $<\sigma^{2}>^{1/2}=$245$\pm$18 km s$^{-1}$ for an $\Omega_m = 0.3$,
 $\Omega_\Lambda=0.7$ universe. This value agrees well with Hoekstra
 et al. (2001) who found $<\sigma^{2}>^{1/2}=$258$\pm$50 km s$^{-1}$
 for an $\Omega_m = 0.2$, $\Omega_\Lambda=0.8$ universe, although
 our result has considerably smaller errors. Our measured velocity dispersion from weak lensing also agrees with the results from a dynamical study of the CNOC2 groups (Carlberg et al. 2001) who found $\sigma\sim$200km s$^{-1}$.

\subsection{Mass-to-Light Ratio}

Gravitational lensing can be used to estimate masses, and hence mass-to-light ratios. The mass
estimate comes directly from the isothermal sphere fit to the
tangential shear data, and the light information comes from the CNOC2
galaxy catalogs (Yee et al. 2000). Each galaxy in the redshift survey
has a measured magnitude and various weights (color, geometric and
 redshift) to account for
incompleteness in the sample. The galaxy group luminosity profile was
calculated by using the magnitudes and weights for each galaxy that
belonged to a galaxy group. The galaxies were placed in radial bins
centered at the group centers, just as was done to measure the lensing
signal. The luminosity of each galaxy was calculated, with a
correction for galaxies fainter than the survey limit. This was done
by employing the
CNOC2 galaxy luminosity function published in Lin et al. (1999) and
using the spectral classification provided in the CNOC2 galaxy
catalogs. The luminosities were not corrected for evolution, but
were k-corrected.
The mass-to-light ratio of the galaxy groups as a function of radius
is plotted in Figure 2. We obtain an integrated mass-to-light ratio to 1.0 h$^{-1}$Mpc of $185\pm28 h$M$_\odot$/L$_{B\odot}$, consistent with the value of $191\pm81 h$M$_\odot$/L$_{B\odot}$ found by
 Hoekstra et al. (2001) using a subset of the groups. The method employed by
  Hoekstra et al. was slightly different in that the mass-to-light
  ratio was estimated by calculating the ratio between the measured shear
  signal and the expected shear derived from the luminosity profile. This
  method requires the assumption that the mass-to-light ratio is
  constant across the groups and was necessary because of the smaller
  data set and low signal-to-noise ratio. As is clear in Figure 2, the $M/L$ is remarkably flat as a function of distance from the group center. This is in contrast to what was found by the dynamical study of the CNOC2 groups, as will be discussed in Section 4. If the M/L is calculated using the NFW mass profile the results are statistically equivalent.

\begin{figure}[h]
\epsscale{.80}
\plotone{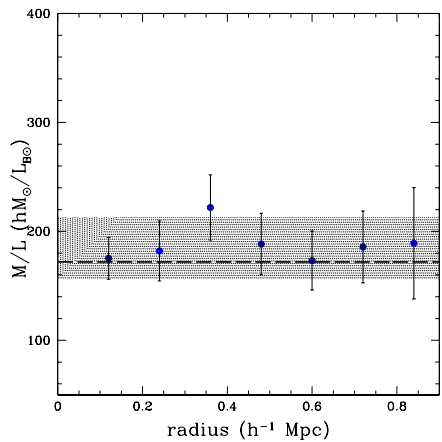}
\caption{The mass-to-light ratio for the entire sample of galaxy groups in radial bins. The average M/L is 185$\pm$28 hM$_\odot$/L$_{B\odot}$. The hatched region represents the 1-$\sigma$ bounds on the mass-to-light ratio, assuming M/L is constant with radius. The heavy dashed line indicates the M/L calculated using all CNOC2 galaxies projected to be close the galaxy group center, as described in the text.}
\end{figure}

It is important to note that the tangential shear signal is sensitive to all matter along the line-of-sight, and as the distance from the group center increases more of the signal is coming from other mass that is correlated with the group (like a 2-halo term in a cross-correlation function). In addition to calculating the M/L of of the galaxy groups using the known group members from the CNOC2 groups catalog, we also calculated the M/L for all galaxies in the CNOC2 galaxy catalog projected to be within a small distance of the group center (1200 km s$^{-1}$ along the line-of-sight). The mass model does not change but the luminosity profile is altered by including more galaxies. The total M/L is lower by 8\%, which is within the 1$\sigma$ errors, and is still flat with distance from the group center. Assuming a constant M/L with distance from the group center, the best fit M/L using this larger sample of galaxies can be observed as the heavy dashed line in Figure 2. 

\subsection{Mass-to-Light Ratios of Rich and Poor Galaxy Groups}
 
We wanted to examine the difference in the shear signal from the rich
and poor groups, and to this end we divided the galaxy group catalog
into two
subsamples. We split the sample by the median dynamical velocity dispersion (190
km s$^{-1}$), although results were similar regardless
of whether the groups were divided by their luminosities or velocity
dispersions. The same source catalog was used to study the two group
subsets. The only difference from the technique outlined in the sections above is that
the input group catalogs have half the number of groups. The resulting tangential and cross shear for the two group
subsets are shown in Figure 3.

\begin{figure}[h]
\epsscale{1.1}
\plottwo{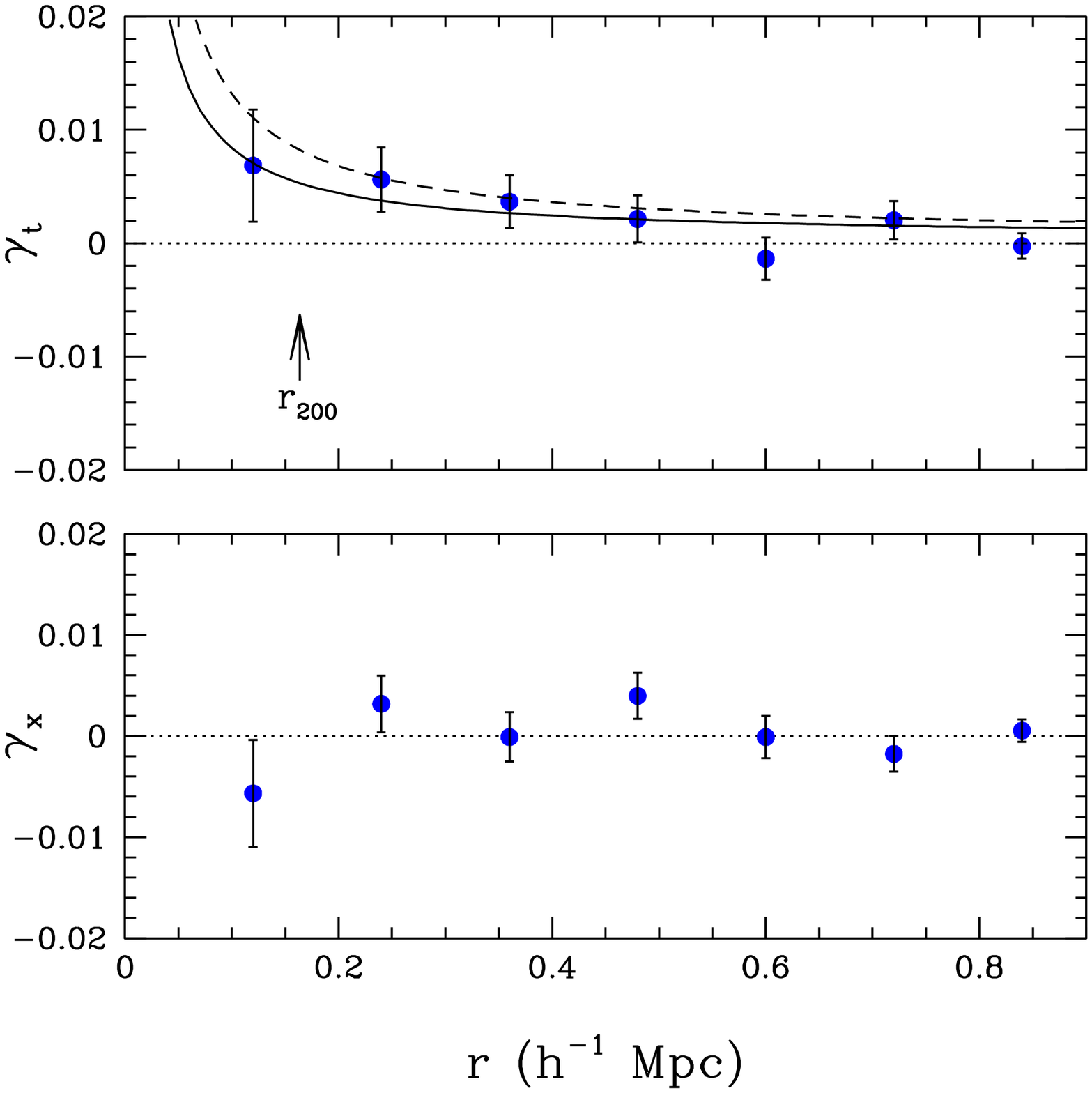}{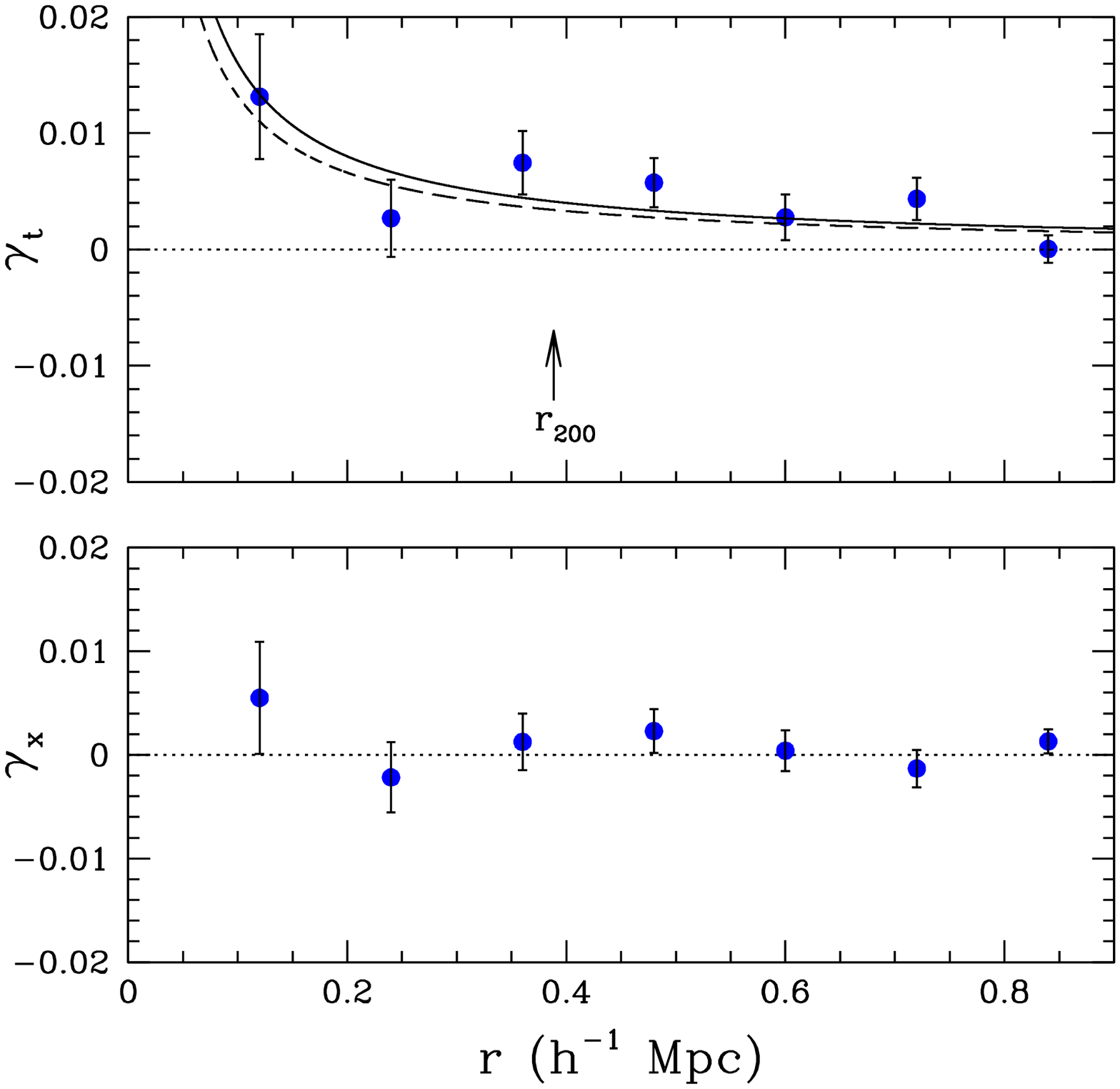}
\caption{The tangential and cross components of shear as in Figure
  1, on the left for the sample of ``poor'' groups, which have dynamical velocity dispersions less than 190 km s$^{-1}$, and on the right for the
  ``rich'' galaxy groups, which have velocity dispersions
  greater than 190 km s$^{-1}$. The best fit isothermal sphere is plotted
  with a solid line in both plots. The best fit velocity dispersion
  is 193$\pm38$ km s$^{-1}$ for the poor groups and 270$\pm$39km s$^{-1}$ for the rich groups. The dashed line is the best
  fit isothermal sphere from Figure 1, for the entire data set. The characteristic $r_{200}$ values (Carlberg et al. 2001) are indicated by the arrow.}
\end{figure}

The results clearly indicate that the lensing signal is dominated by
the larger groups. A few of the galaxy groups identified in the CNOC2
fields had measured velocity dispersions in excess of 500 km s$^{-1}$. To be certain that the tangential shear signal measured was not
coming solely from these groups we measured the tangential shear around only
those groups with velocity dispersions greater than 500 km s$^{-1}$. In
addition, we also repeated the measurement of the tangential shear around all galaxies
except those with velocity dispersions greater than 500 km s$^{-1}$. The
results indicated that there is a
substantial signal coming from the most massive galaxy groups (small
clusters) but that the tangential shear profile is not dominated by
these for the sample as a whole. After the massive groups are
removed an isothermal tangential shear profile with an Einstein radius
of roughly 0.8$''$ remains.

The shear profiles for the two subsamples were fit with isothermal
spheres and their mass-to-light ratios were estimated. The mean velocity dispersion of the ``poor'' groups is $<\sigma^{2}>^{1/2}=$193$\pm$38 km s$^{-1}$, while the ``rich'' groups have a velocity dispersion of $<\sigma^{2}>^{1/2}=$270$\pm$39 km s$^{-1}$. The mass-to-light ratios of the ``rich'' and ``poor'' galaxy groups are
flat with radius, as can be seen in Figure 4. The weighted mean mass-to-light ratio of the ``poor''
groups is 134$\pm$26 hM$_{\odot}$/L$_{B\odot}$, while
the mass-to-light ratio of the ``rich'' groups is 278$\pm$42 hM$_{\odot}$/L$_{B\odot}$. Once again, the M/Ls were also calculating using a catalog of galaxies projected to be close to the group center and the best fit results are shown with the heavy dashed lines in Figure 4.

\section{Discussion}

Weak lensing is a powerful tool for understanding the ensemble-averaged
 properties for a sample of objects, but can not tell us about the
 properties of an individual galaxy group. It is of great interest to
 compare the results we obtained for our sample of CNOC2 galaxy
 groups to those obtained by Carlberg et al.(2001), using dynamical
 methods on the full sample of CNOC2 galaxy groups. In Figure 5 we have plotted the mass-to-light ratio of
 galaxy groups as a function of radius from the lensing, as well as the best
 fit curve for the dynamical data. Note that the M/L ratio is plotted in units of $r_{200}$. Our values for the $r_{200}$ for each group came from the Carlberg et al. galaxy group catalog. We repeated the compete analysis of the shear and of the luminosity based in units of $r_{200}$. We do not observe the steep increase in M/L that was observed with the dynamical methods, and our data can be well-fit with a straight line with no slope. The M/L calculated from the dynamical data is dependent on the orbits of the galaxies in the groups, but their result of a rising M/L is robust for all reasonable orbits. The lensing-,and dynamically-derived properties of these CNOC2 groups are outlined in Table 2.

\begin{figure}[h]
\epsscale{.80}
\plotone{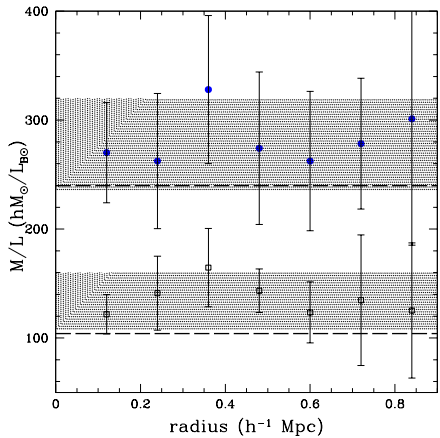}
\caption{The mass-to-light ratio for subsamples of galaxy groups in radial bins. Symbols are as in Figure 2. The mass-to-light ratio of the poor galaxy groups (open
  squares) and rich galaxy groups (filled circles) as a function of
  the distance from the group center. The average M/L of the rich groups is 278$\pm$42 hM$_\odot$/L$_{B\odot}$and
  134$\pm$26 hM$_\odot$/L$_{B\odot}$ for the poor groups. There is a
  clear offset in the mass-to-light ratio for the two subsamples as can
  be seen by comparing the two hatched regions.}
\end{figure}

\begin{table*}
\begin{center}
\caption{Weak lensing and dynamical properties of galaxy groups\label{tbl-2}}
\begin{tabular}{cccccccccc}
\tableline
Sample & mean N$_{gal}$ & $<\sigma>_{lensing}$ & $<\sigma>_{dyn}$\tablenotemark{a} & $<$M/L$>_{B}$ & median redshift\\
& per group & km s$^{-1}$ & km s$^{-1}$ & hM$_{\odot}$/L$_{\odot}$ &\\
\tableline

all groups & 3.9 & 245$\pm18$ & 219$\pm$10 & 185$\pm$28 & 0.323\\
rich groups & 4.2 & 270$\pm38$ & 311$\pm$13 & 278$\pm$42 & 0.360 \\
poor groups & 3.6 & 198$\pm38$ & 127$\pm$4 & 134$\pm$26 & 0.303\\

\tableline
\end{tabular}
\tablenotetext{a}{Carlberg et al. 2001}
\end{center}
\end{table*} 

It is particularly interesting to plot our two values for the
mass-to-light ratio of galaxy groups on the mass sequence, and compare
our results to previous measurements. This can be seen in Figure 6
where B-band mass-to-light ratios for galaxy groups and clusters are
plotted. The curve is from Marinoni \& Hudson (2002), who
estimated the mass-to-light ratio by comparing the mass function
from Press-Schechter theory with their measured luminosity
function of virialized systems. Our data follow the trend of rising mass-to-light ratio with
mass. We are in approximate agreement with the global mass-to-light
ratio found on similar scales by
Carlberg et al. (2001), Tucker et al. (2000) and Eke et al. (2004),
although some of these studies are at a lower redshift.

\begin{figure}[h]
\epsscale{.80}
\plotone{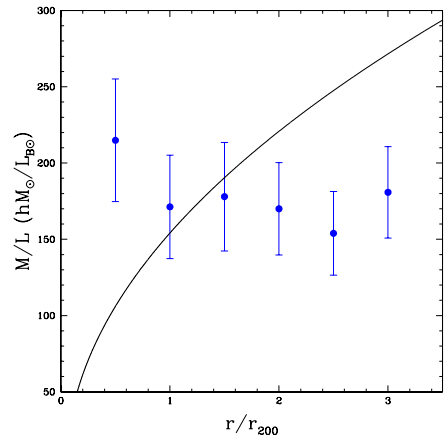}
\caption{The mass-to-light ratio of the galaxy groups in radial bins
  in units of r$_{200}$. The line is fit to the dynamical results of
  Carlberg et al. (2001). We do not observe the steep rise in mass-to-light ratio with radius. Note that the Carlberg et al. results are for a
  3-dimensional mass-to-light ratio while the data points are in
  projection. }
\end{figure}

The halo mass of approximately 10$^{13}$M$_{\odot}$, hosting typically
3 L* galaxies (Marinoni \& Hudson 2002), appears to be a critical
scale, at which the mass-to-light ratio is increasing dramatically as
a function of mass. This could indicate the transition from the
actively star-forming field environment to the
passively-evolving-galaxy-dominated cluster regime. Note that the M/L
of the rich galaxy groups appears to be comparable to that found in
massive galaxy clusters (Carlberg et al. 1997), as seen in Figure 6,
so there is presumably little change in M/L on more massive scales.

A rise in M/L on these scales has been suggested previously, from
analysis based directly on dynamical studies of groups (Marinoni \&
Hudson 2002; van den Bosch, Yang and Mo 2003; Eke et al. 2004) and from
semi-analytic models (for example Benson et al. 2000), see also discussion in Dekel
\& Birnboim (2004). The results of this
paper are then consistent with previous studies, although this weak
lensing result suggests a somewhat steeper increase than had been
found previously.

\begin{figure}[h]
\epsscale{.80}
\plotone{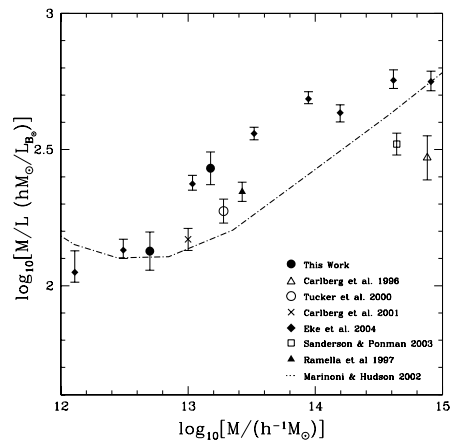}
\caption{The mass-to-light ratio as a function of mass for samples of
  groups and clusters from the literature. The curve (Marinoni \& Hudson, 2002) is generated
  by comparing the mass function from Press-Schechter theory for a
  $\Lambda$CDM universe with an observed luminosity function. Note
  that the different samples span a range of redshifts. For
  example, our median redshift is 0.33 while the 2PIGG data (Eke et al. 2004) median
  redshift is roughly 0.1.}
\end{figure}

From the average mass-to-light ratio for our entire sample of galaxy
groups it
is possible to naively estimate $\Omega_m$ using the method outlined in Carlberg
et al. (1997). By combining M/L with luminosity density it is possible
to  estimate
the matter density of the Universe.

\begin{equation}
\Omega_m=\frac{\rho_m}{\rho_c}=\frac{\rho_L}{\rho_c}\frac{M}{L}
\end{equation}
Using a mass-to-light ratio of 185$\pm28$
hM$_{\odot}$/L$_{R\odot}$ (converted to hM$_{\odot}$/L$_{R_{AB}\odot}$),
and the luminosity density for CNOC2 galaxies found in Lin et
al. (1999) we obtain $\Omega_m$=0.22$\pm$0.06. This is a valid estimate of $\Omega_m$ only if galaxy groups dominate the mass and luminosity of the Universe. In order to properly
calculate $\Omega_m$, it is necessary to know the mass-to-light
function for a
wide range of masses, extending to single galaxies and rich clusters.

\section{Conclusions}

We have detected a significant weak lensing signal for a sample of 116
intermediate redshift galaxy groups. From the lensing signal we
estimate that galaxy groups have a mean M/L of 185$\pm$28
hM$_\odot$/L$_{B\odot}$ within 1 h$^{-1}$Mpc, and that this M/L is constant as the distance from the group center increases. When the sample is split into subsets of rich and poor galaxy groups, there is a clear offset in the
mass-to-light ratios of the two subsets. The increase in the M/L
as a function of mass is in general agreement with other results, but
is detected here for the first time using weak lensing in the galaxy
group mass regime.

This analysis indicates that a weak lensing signal can indeed be
measured from galaxy groups. Clearly, a larger sample with well determined dynamical
properties would be ideal for this sort of study. The structure of the
dark matter
halos of galaxy groups are still poorly understood. By combining
this group lensing result with galaxy-galaxy lensing it should be
possible to determine the size and extent of galaxy group dark
matter halos, which will aid significantly in our understanding of
structure in the Universe and the nature of dark matter.

\acknowledgments

MJH acknowledges support from NSERC and a Premier's Research Excellence 
Award.

\clearpage

\end{document}